\documentclass[prl,twocolumn,superscriptaddress,showpacs,preprintnumbers,amsmath,amssymb]{revtex4-1}

\usepackage{graphicx}% Include figure files
\usepackage{dcolumn}% Align table columns on decimal point
\usepackage{bm}% bold math
\usepackage{nicefrac}% fractions like 1/2
\usepackage{color,ulem}% color and ulem
\usepackage[dvipdfmx, colorlinks=True,citecolor=blue,linkcolor=blue,urlcolor=blue]{hyperref}

\begin{document}

%\preprint{APS/PRL}

\title{$\mathbf{q}=\mathbf{0}$ long-range magnetic order in centennialite CaCu$_3$(OD)$_6$Cl$_2$$\cdot$0.6D$_2$O: A spin-1/2 perfect kagome antiferromagnet with $J_1$-$J_2$-$J_d$}% Force line breaks with \\

\author{K.~Iida}\email{k\_iida@cross.or.jp}
\affiliation{Neutron Science and Technology Center, Comprehensive Research Organization for Science and Society (CROSS), Tokai, Ibaraki 319-1106, Japan}

\author{H.~K.~Yoshida}
\affiliation{Department of Physics, Faculty of Science, Hokkaido University, Sapporo, Hokkaido 060-0810, Japan}

\author{A.~Nakao}
\affiliation{Neutron Science and Technology Center, Comprehensive Research Organization for Science and Society (CROSS), Tokai, Ibaraki 319-1106, Japan}

\author{H.~O.~Jeschke}
\affiliation{Research Institute for Interdisciplinary Science, Okayama University, Okayama 700-8530, Japan}

\author{Y.~Iqbal}
\affiliation{Department of Physics, Indian Institute of Technology Madras, Chennai 600036, India}

\author{K.~Nakajima}
\affiliation{J-PARC Center, Japan Atomic Energy Agency (JAEA), Tokai, Ibaraki 319-1195, Japan}

\author{S.~Ohira-Kawamura}
\affiliation{J-PARC Center, Japan Atomic Energy Agency (JAEA), Tokai, Ibaraki 319-1195, Japan}

\author{K.~Munakata}
\affiliation{Neutron Science and Technology Center, Comprehensive Research Organization for Science and Society (CROSS), Tokai, Ibaraki 319-1106, Japan}

\author{Y.~Inamura}
\affiliation{J-PARC Center, Japan Atomic Energy Agency (JAEA), Tokai, Ibaraki 319-1195, Japan}

\author{N.~Murai}
\affiliation{J-PARC Center, Japan Atomic Energy Agency (JAEA), Tokai, Ibaraki 319-1195, Japan}

\author{M.~Ishikado}
\affiliation{Neutron Science and Technology Center, Comprehensive Research Organization for Science and Society (CROSS), Tokai, Ibaraki 319-1106, Japan}

\author{R.~Kumai}
\affiliation{Institute of Materials Structure Science, High Energy Accelerator Research Organization (KEK), Tsukuba, Ibaraki 305-0801, Japan}
\affiliation{Department of Materials Structure Science, The Graduate University for Advanced Studies, Tsukuba, Ibaraki 305-0801, Japan}

\author{T.~Okada}
\affiliation{Department of Physics, Faculty of Science, Hokkaido University, Sapporo, Hokkaido 060-0810, Japan}

\author{M.~Oda}
\affiliation{Department of Physics, Faculty of Science, Hokkaido University, Sapporo, Hokkaido 060-0810, Japan}

\author{K.~Kakurai}
\affiliation{Neutron Science and Technology Center, Comprehensive Research Organization for Science and Society (CROSS), Tokai, Ibaraki 319-1106, Japan}

\author{M.~Matsuda}
\affiliation{Neutron Scattering Division, Oak Ridge National Laboratory, Oak Ridge, Tennessee 37831, USA}

\date{\today}% It is always \today, today,
             %  but any date may be explicitly specified

\begin{abstract}
Crystal and magnetic structures of the mineral centennialite CaCu$_3$(OH)$_6$Cl$_2\cdot0.6$H$_2$O are investigated by means of synchrotron x-ray diffraction and neutron diffraction measurements complemented by density functional theory (DFT) and pseudofermion functional renormalization group (PFFRG) calculations.
CaCu$_3$(OH)$_6$Cl$_2\cdot0.6$H$_2$O crystallizes in the $P\bar{3}m1$ space group and Cu$^{2+}$ ions form a geometrically perfect kagome network with antiferromagnetic $J_1$.
No intersite disorder between Cu$^{2+}$ and Ca$^{2+}$ ions is detected.
CaCu$_3$(OH)$_6$Cl$_2\cdot0.6$H$_2$O enters a magnetic long-range ordered state below $T_\text{N}=7.2$~K, and the $\mathbf{q}=\mathbf{0}$ magnetic structure with negative vector spin chirality is obtained.
The ordered moment at 0.3~K is suppressed to $0.58(2)\mu_\text{B}$.
Our DFT calculations indicate the presence of antiferromagnetic $J_2$ and ferromagnetic $J_d$ superexchange couplings of a strength which places the system at the crossroads of three magnetic orders (at the classical level) and a spin-$\frac{1}{2}$ PFFRG analysis shows a dominance of $\mathbf{q}=\mathbf{0}$ type magnetic correlations, consistent with and indicating proximity to the observed $\mathbf{q}=\mathbf{0}$ spin structure.
The results suggest that this material is located close to a quantum critical point and is a good realization of a $J_1$-$J_2$-$J_d$ kagome antiferromagnet.\end{abstract}

%\pacs{74.70.Pq, 78.70.Nx}% PACS, the Physics and Astronomy
                             % Classification Scheme.
%\keywords{Suggested keywords}%Use showkeys class option if keyword
                              %display desired
\maketitle

The search for quantum spin liquid (QSL) has been pursued intensively after the initial idea of a resonating valence bond was proposed~\cite{RVB}.
The spin-1/2 kagome lattice with the nearest neighbor antiferromagnetic $J_1$ is the prime candidate for hosting QSLs~\cite{Review1,Review2}.
Several types have been proposed theoretically, but consensus for example concerning gapless or gapped excitations has yet to be reached.
Herbertsmithite, ZnCu$_3$(OH)$_6$Cl$_2$, is the most investigated material in this context~\cite{ZnHerber_Shores_2005,ZnHerber_Mendels_2010,ZnHerber_Norman_2016} and shows QSL behavior down to $50$~mK~\cite{ZnHerber_Mendels_2007,ZnHerber_Han_2012}.
There is nonnegligible intersite disorder between Cu$^{2+}$ and Zn$^{2+}$, which may hinder the true ground state~\cite{ZnHerber_Lee_2007,ZnHerber_Olariu_2008}.
On the other hand, YCu$_3$(OH)$_6$Cl$_2$~\cite{YCu3_Zorko_uSR_2019,YCu3_Zorko_2019,YCu3_Barthelemy_2019} and CdCu$_3$(OH)$_6$(NO$_3$)$_2\cdot$H$_2$O~\cite{Cdkagome_Okuma_2017} are disorder free spin-1/2 perfect kagome lattice antiferromagnets and exhibit $\mathbf{q}=\mathbf{0}$ long-range magnetic order owing to the Dzyaloshinskii-Moriya (DM) interaction~\cite{DM_Cepas_2008}.

\begin{figure}[t]
\includegraphics[width=\linewidth]{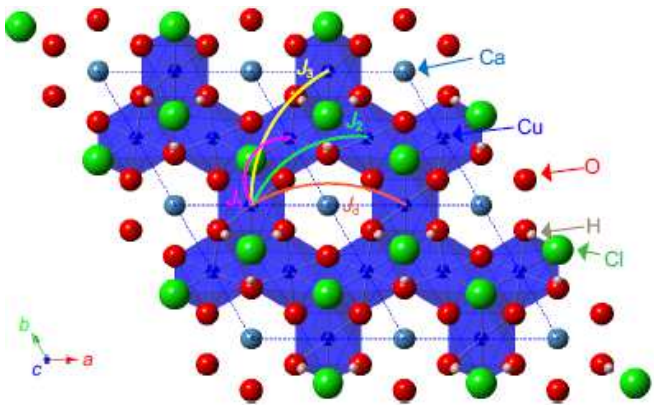}
\centering
\caption{
Crystal structure of CaCu$_3$(OH)$_6$Cl$_2\cdot0.6$H$_2$O.
The exchange couplings $J_1$, $J_2$, $J_3$, and $J_d$ are also illustrated.
Dotted lines represent the structural unit cell.}\label{Fig:Structure}
\end{figure}

There is an alternative attempt to search for a QSL state in kagome compounds, in which competing interactions such as $J_2$ and $J_d$ across the hexagon (see Fig.~\ref{Fig:Structure}) are introduced.
Kapellasite, which is a polymorph of herbertsmithite, is the prototype compound of the $J_1$-$J_2$-$J_d$ kagome lattice magnet~\cite{Kapellasite_Janson_2008,ZnKape_Colman_2008,ZnKape_Fak_2012,ZnKape_Kermarrec_2014}.
Zn$^{2+}$ in herbertsmithite is located between the kagome planes, whereas Zn$^{2+}$ in kapellasite is located at the center of a hexagon of a kagome layer, resulting in the finite strength of $J_2$ and $J_d$~\cite{Kapellasite_Janson_2008}.
Eventually, the kapellasite family $A$Cu$_3$(OH)$_6$Cl$_2$ where $A$ is Zn or Mg provides a rich phase diagram as a function of $J_2$ and $J_d$~\cite{ClassicalEnergy,HTSE,CdKapellasite_Iqbal_2015}.
Both ZnCu$_3$(OH)$_6$Cl$_2$ and MgCu$_3$(OH)$_6$Cl$_2$ (known as haydeeite) possess ferromagnetic $J_1$~\cite{ZnKape_Fak_2012,MgKape_Boldrin_2015}.
Haydeeite undergoes ferromagnetic order below $T_\text{C}=4.2$~K~\cite{MgKape_Boldrin_2015}, while kapellasite shows QSL behavior down to 20~mK~\cite{ZnKape_Fak_2012}.
Meanwhile, both ZnCu$_3$(OH)$_6$Cl$_2$ and MgCu$_3$(OH)$_6$Cl$_2$ have intersite disorder between $A^{2+}$ and Cu$^{2+}$~\cite{ZnKape_Kermarrec_2014,MgKape_Colman_2010}.
CdCu$_3$(OH)$_6$Cl$_2$ was theoretically proposed to carry antiferromagnetic $J_1$~\cite{CdKapellasite_Iqbal_2015} but it has not been synthesized yet.
Instead, CdCu$_3$(OH)$_6$(NO$_3$)$_2\cdot$H$_2$O (Cd kapellasite) is reported to exhibit antiferromagnetic $J_1$ but is discussed in terms of $J_1$ and DM interactions~\cite{Cdkagome_Okuma_2017}.

Recently, CaCu(OH)$_6$Cl$_2\cdot0.6$H$_2$O (centennialite), a sister mineral of kapellasite and haydeeite, was discovered~\cite{CaCu3_2014,CaCu3_Crichton_2017,CaCu3_Sun_2016}.
Centennialite has an ideal kagome network and Ca$^{2+}$ ion is located at the center of the hexagon as described in Fig.~\ref{Fig:Structure}.
Remarkably, large single crystals are available for centennialite~\cite{CaCu3_Yoshida_2017}.
The magnetic susceptibilities exhibit negative Curie-Weiss temperature ($\Theta_\text{CW}$), indicating that the dominant interaction is antiferromagnetic $J_1$~\cite{CaCu3_Yoshida_2017}.
Furthermore, the high temperature series expansion (HTSE) reveals the existence of $J_2$ and $J_d$ as in kapellasite~\cite{ZnKape_Fak_2012} and haydeeite~\cite{MgKape_Boldrin_2015}.
Since kapellasite and haydeeite have ferromagnetic $J_1$~\cite{ZnKape_Fak_2012,MgKape_Boldrin_2015}, centennialite is considered to be an ideal compound for investigating the $J_1$-$J_2$-$J_d$ kagome lattice with antiferromagnetic $J_1$~\cite{CaCu3_Yoshida_2017}.
The heat capacity~\cite{CaCu3_Yoshida_2017} and nuclear magnetic resonance~\cite{CaCu3_Ihara_2017} measurements show that centennialite undergoes long-range magnetic ordering below $T_\text{N}=7.2$~K.
However, detailed crystal and magnetic structures of centennialite have not been reported yet by diffraction techniques.
In this paper, we investigated in detail the crystal and magnetic structures by synchrotron x-ray and neutron diffraction measurements using single crystals.
In addition, we also performed density functional theory (DFT) calculations to theoretically evaluate the exchange couplings in centennialite, and subsequently employed the pseudofermion functional renormalization group (PFFRG) method~\cite{Reuther-2010} to reveal the momentum resolved magnetic susceptibility profile for the DFT Hamiltonian. We demonstrate that CaCu(OH)$_6$Cl$_2\cdot0.6$H$_2$O is the first $J_1$-$J_2$-$J_d$ kagome lattice antiferromagnet and hence offer the new playground for geometrical frustration.

Single crystals of centennialite and deuterated centennialite were grown by a hydrothermal reaction~\cite{CaCu3_Yoshida_2017}.
Precise crystal structure determination of centennialite was performed using the imagingplate x-ray diffractometer BL-8A installed at Photon Factory (PF).
Incident x-ray energy of 18~keV was used.
Neutron diffraction measurements on deuterated single crystals of centennialite were carried out using the time-of-flight (TOF) chopper spectrometers AMATERAS~\cite{AMATERAS,Utsusemi} and 4SEASONS~\cite{Utsusemi,4SEASONS_1,4SEASONS_2,SM}, and the TOF neutron diffractometer SENJU~\cite{SENJU} installed at Materials and Life Science Experimental Facility (MLF), Japan Proton Accelerator Research Complex (J-PARC).
The incident energy of neutrons $E_i=5.24$~meV with the energy resolution of 0.16~meV (full width at half maximum) at the elastic channel was used at AMATERAS.
A neutron wavelength of $0.4\sim4.4$~$\text{\AA}$ was used at SENJU.

We perform density functional theory calculations based on the all electron full potential local orbital (FPLO) basis~\cite{Koepernik1999} with generalized gradient approximation (GGA) exchange correlation functional~\cite{Perdew1996}.
We use a GGA$+U$ correction~\cite{Liechtenstein1995} to account for the strong electron correlations on the Cu$^{2+}$ $3d$ orbitals.
We deal with the small disorder of the Ca and water molecule positions by fixing Ca in the Cu plane and by simplifying to 1/2 water molecules per formula unit.
We apply the energy mapping technique~\cite{Guterding2016,Iqbal2017} to extract the Heisenberg Hamiltonian parameters of centennialite~\cite{SM}. To gain insight into the magnetic behavior of CaCu(OH)$_6$Cl$_2\cdot0.6$H$_2$O, we employ the theoretical framework of PFFRG~\cite{Reuther-2010} to compute the real part of the static and momentum-resolved spin susceptibility expected from this set of interactions.

Synchrotron x-ray diffraction measurements on single-crystal CaCu$_3$(OH)$_6$Cl$_2\cdot0.6$H$_2$O are performed to determine the crystal structure below and above $T_\text{N}$, and the refined crystal structure parameters are summarized in the Supplemental Material~\cite{SM}.
As kapellasite~\cite{ZnKape_Colman_2008} and haydeeite~\cite{MgKape_Colman_2010}, CaCu$_3$(OH)$_6$Cl$_2\cdot0.6$H$_2$O crystallizes in the $P\bar{3}m1$ space group, which persists below $T_\text{N}$.
The Cu$^{2+}$ sites form a structurally perfect kagome lattice (Fig.~\ref{Fig:Structure}), separated by $5.74$~$\text{\AA}$ along the $c$ axis, giving rise to good two dimensionality.
The mixing amount between Cu$^{2+}$ and Ca$^{2+}$ ions is zero within errors because of the large difference of their ionic radii~\cite{CaCu3_Sun_2016,CaCu3_Yoshida_2017}.
The large ionic radius induces disorder of the Ca site along the $c$ axis, and the average position is the center of the hexagon of the kagome plane.

\begin{figure}[t]
\includegraphics[width=\linewidth]{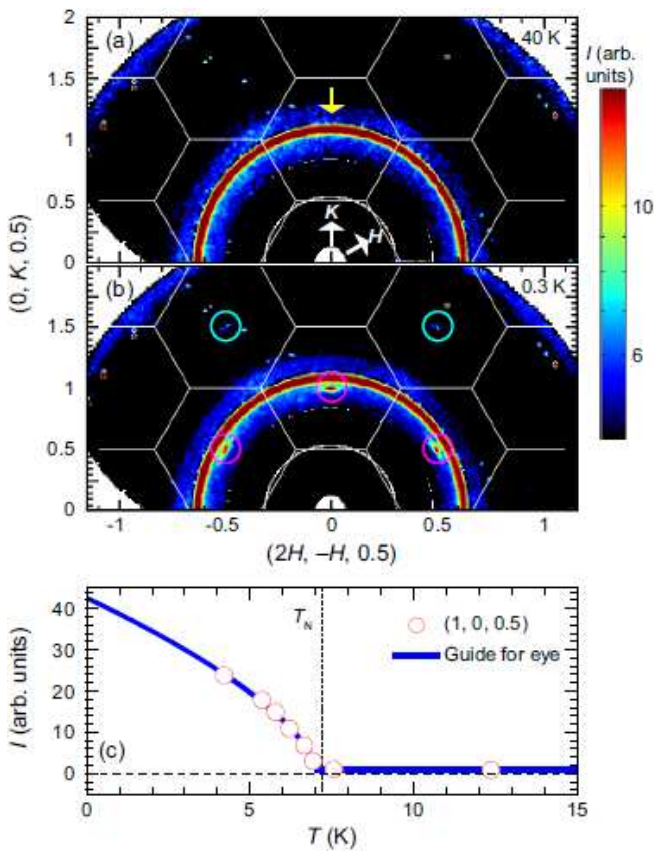}
\centering
\caption{
(a, b) Reciprocal-space neutron scattering intensity maps of CaCu$_3$(OD)$_6$Cl$_2\cdot0.6$D$_2$O at (a) 40~K and (b) 0.3~K measured at AMATERAS.
Energy transfer ($\hbar\omega$) is averaged over $[-0.15, 0.15]$~meV.
Solid lines represent the two-dimensional structural Brillouin zones.
Pink and blue circles in panel (b) represent, respectively, the magnetic reflections at $\mathbf{Q}=(1,0,0.5)$ and $(1,1,0.5)$ with their equivalent positions.
Single crystals are fixed by the hydrogen-free grease which gives the ring-shape background at $Q=1.3$~$\text{\AA}^{-1}$ [yellow arrow in panel (a)] as described in detail in the Supplemental Material~\cite{SM}.
(c) Temperature dependence of the neutron scattering intensity at $\mathbf{Q}=(1,0,0.5)$ measured at SENJU.
The dotted line represents $T_\text{N}=7.2$~K determined by the heat capacity measurements~\cite{CaCu3_Yoshida_2017}.
}\label{Fig:Diffraction}
\end{figure}

To understand the magnetic ground state of CaCu$_3$(OD)$_6$Cl$_2\cdot0.6$D$_2$O, elastic neutron scattering measurements were performed at AMATERAS.
Figures~\ref{Fig:Diffraction}(a) and \ref{Fig:Diffraction}(b) depict the reciprocal-space neutron scattering intensity maps of CaCu$_3$(OD)$_6$Cl$_2\cdot0.6$D$_2$O at $T=40$ and 0.3~K in the $(HK0.5)$ plane.
Evolution of several peaks of $10\frac{1}{2}$ and $11\frac{1}{2}$ with their equivalent positions can be seen at 0.3~K [see circles in Fig.~\ref{Fig:Diffraction}(b)].
The peak at $\mathbf{Q}=\left(0,1,\frac{1}{2}\right)$ has the resolution-limited peak width~\cite{SM}, indicative of the long-range magnetic order.
The temperature dependence of the neutron scattering intensity at $\mathbf{Q}=(1,0,0.5)$ momentum transfer is shown in Fig.~\ref{Fig:Diffraction}(c).
Clearly, the peak at $\mathbf{Q}=(1,0,0.5)$ develops below $T_\text{N}$, revealing the magnetic origin.
In addition, magnetic reflections are also observed at higher $\mathbf{Q}$~\cite{SM}.

\begin{figure}[t]
\includegraphics[width=\linewidth]{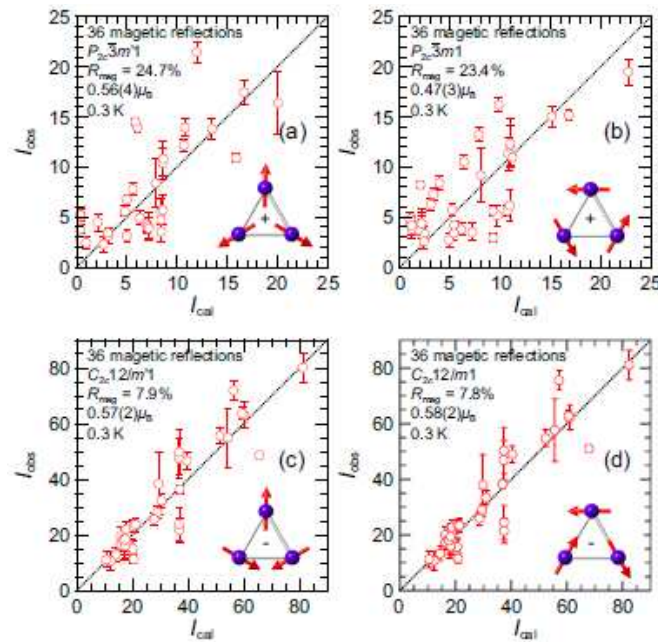}
\centering
\caption{
Observed and calculated magnetic structure factors of CaCu$_3$(OD)$_6$Cl$_2\cdot0.6$D$_2$O.
Experimental data are measured at 0.3~K using SENJU~\cite{SM}.
Magnetic structures with (a, b) positive vector spin chirality and (c, d) negative vector spin chirality are refined.
}\label{Fig:Fitting}
\end{figure}

The observed magnetic reflections are located at the center of the two-dimensional structural Brillouin zones [Fig.~\ref{Fig:Diffraction}(b)].
This result indicates the $\mathbf{q}=\mathbf{0}$ magnetic structure, and the corresponding magnetic propagation vector is $\mathbf{k}=(0,0,0.5)$.
There are, however, several possibilities of the spin textures for the magnetic propagation vector~\cite{YCu3_Zorko_2019,Cdkagome_Okuma_2017}, some of which are shown in the insets of Fig.~\ref{Fig:Fitting}.

For unambiguous determination of the magnetic structure of CaCu$_3$(OD)$_6$Cl$_2\cdot0.6$D$_2$O, we further perform detailed neutron diffraction measurements at SENJU using a deuterated single crystal with the dimensions of $1.2\times1.2\times0.3$~mm$^3$.
Magnetic structure analysis was performed using the JANA2006 software~\cite{JANA}.
Considering the crystal symmetry $P\bar{3}m1$ with the magnetic propagation vector of $\mathbf{k}=(0,0,0.5)$, the candidates for the magnetic space group of CaCu$_3$(OD)$_6$Cl$_2\cdot0.6$D$_2$O are $P_{2c}\bar{3}m'1$, $P_{2c}\bar{3}m1$, $C_{2c}12/m'1$, $C_{2c}12/m1$, $P\bar{1}$, and $P1$.
We refined the neutron diffraction data using trigonal and monoclinic symmetries among them.
For the $C$-monoclinic models with the $2a\times(a+b)\times c$ unit cell, the 120$^\circ$ spin structure with the same moment sizes for two independent Cu sites [see the insets of Fig.~\ref{Fig:Fitting}(c) and \ref{Fig:Fitting}(d)] and three twinned domains owing to a trigonal-to-monoclinic modulation are considered.
In all models, we assumed that the magnetic moments lie in the kagome plane.
We refined the four models using the 36 magnetic reflections [$I>3\sigma(I)$].
Refinement results and the reliability factors are described in Fig.~\ref{Fig:Fitting}.
Apparently, the magnetic structures belonging to $C_{2c}12/m'1$ and $C_{2c}12/m1$ with the negative vector spin chirality are favored over $P_{2c}\bar{3}m'1$ and $P_{2c}\bar{3}m1$ with the positive vector spin chirality.
On the other hand, the present data set cannot distinguish the $C_{2c}12/m'1$ and $C_{2c}12/m1$ models because of the domain contributions.
The above assumption that the magnetic moments lie in the kagome plane is justified by the discussion in Refs.~[\onlinecite{YCu3_Zorko_2019,DM2002}]; the $\mathbf{q}=\mathbf{0}$ spin structure with the negative vector spin chirality energetically favors the coplanar spin structure over the noncoplanar canted spin structure within the realistic magnetic anisotropy in contrast to the case with the positive vector spin chirality as realized in the iron jarosite~\cite{jarosite}.
In addition, the spin correlations with the negative spin chirality is consistent with the recent nuclear magnetic resonance measurements on single-crystal CaCu$_3$(OH)$_6$Cl$_2\cdot0.6$H$_2$O~\cite{CaCu3_Ihara_2020}.
Therefore, the $\mathbf{q}=\mathbf{0}$ spin texture with the negative vector spin chirality is realized in CaCu$_3$(OD)$_6$Cl$_2\cdot0.6$D$_2$O as in YCu$_3$(OH)$_6$Cl$_2$~\cite{YCu3_Zorko_2019} and CdCu$_3$(OH)$_6$(NO$_3$)$_2\cdot$H$_2$O~\cite{Cdkagome_Okuma_2017}.

The obtained ordered moment of centennialite at 0.3~K is $0.58(2)\mu_\text{B}$ which is reduced about $40\%$ from the ideal value for the spin-1/2 system.
The ordered moment in the $J_1$-only perfect kagome lattice antiferromagnet YCu$_3$(OH)$_6$Cl$_2$ is $0.42(2)\mu_\text{B}$ (reduced about $60\%$)~\cite{YCu3_Zorko_2019}.
Thus, $J_2$ and $J_d$ in centennialite suppress the quantum fluctuations compared to YCu$_3$(OH)$_6$Cl$_2$.
On the other hand, the frustration parameter $\left|\Theta_\text{CW}\right|/T_\text{N}$ in CaCu$_3$(OH)$_6$Cl$_2\cdot0.6$H$_2$O is 7.8~\cite{CaCu3_Yoshida_2017} which is higher than 6.6 for YCu$_3$(OH)$_6$Cl$_2$~\cite{YCu3_Zorko_2019,YCu3_Sun_2016}.
This result indicates that the geometrical frustration owing to the kagome network persists in CaCu$_3$(OH)$_6$Cl$_2\cdot0.6$H$_2$O even in the presence of finite $J_2$ and $J_d$.

\begin{table}[t]
\begin{center}
\caption{
Exchange interactions of CaCu$_3$(OH)$_6$Cl$_2\cdot0.6$H$_2$O calculated within GGA$+U$.
See Fig.~\ref{Fig:Structure} for the definition of $J_1$, $J_2$, $J_3$, and $J_d$.
The Curie-Weiss temperature is calculated by the equation $\Theta_\text{CW}=-\frac{2}{3}S (S + 1) \big(2 J_1 + 2 J_2 + 2 J_3 + J_d\big)$~\cite{SM}.
}\label{Tab:DFT}
\begin{tabular}{llllll}
\hline
\hline
$U$ (eV)\ \ & \ $J_1$ (K)\ \ \ \ & $J_2$ (K)\ \ \ \ & $J_{3}$ (K)\ \ \ & $J_{d}$ (K)\ \ \ \ \ \ & $\Theta_\text{CW}$ (K)\\
\hline
6.0 &\ 74(2) & 3.4(1.9) & 0.0(2.2) & $-2.3$(3.7) & $-76.3$\\
7.0 &\ 64(2) & 2.8(1.6) & 0.0(1.8) & $-2.0$(3.1) & $-65.8$\\
8.0 &\ 55(2) & 2.2(1.3) & 0.0(1.5) & $-1.7$(2.5) & $-56.4$\\
\hline
\hline
\end{tabular}
\end{center}
\end{table}

\begin{figure}[t]
\includegraphics[width=\linewidth]{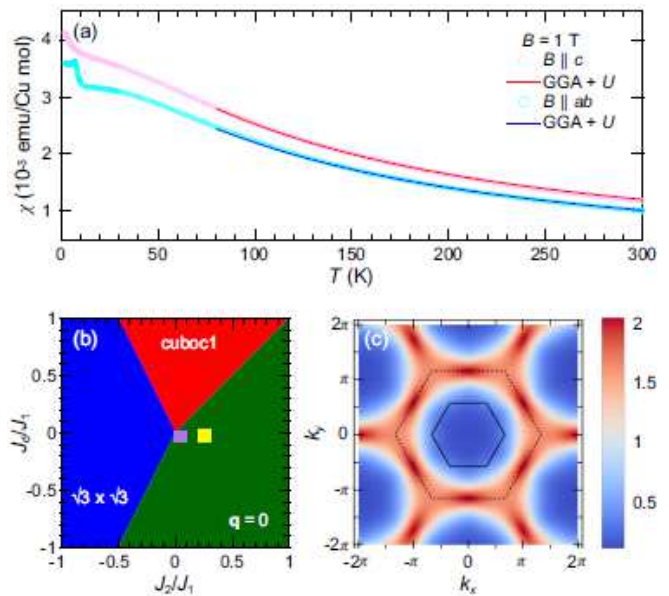}
\centering
\caption{
(a) Temperature dependencies of the magnetic susceptibilities of CaCu$_3$(OH)$_6$Cl$_2\cdot0.6$H$_2$O under the external magnetic field $B=1$~T.
Solid lines are the HTSE results~\cite{HTSE} using $J_1$, $J_2$, and $J_d$ obtained by the GGA$+U$ calculations in Table~\ref{Tab:DFT}.
The experimental data is taken from Ref.~[\onlinecite{CaCu3_Yoshida_2017}].
(b) Classical phase diagram of the $J_1$-$J_2$-$J_d$ kagome-lattice magnet for antiferromagnetic $J_1$~\cite{ClassicalEnergy,HTSE}.
The cuboc1 state is the twelve sublattice noncoplanar spin structures~\cite{cuboc}.
Purple and yellow solid squares represent the GGA$+U$ result and the HTSE fitting result~\cite{SM} for CaCu$_3$(OH)$_6$Cl$_2\cdot0.6$H$_2$O. (c) Momentum resolved spin susceptibility (in units of $1/J_{1}$) profile obtained using PFFRG evaluated at the lowest simulated temperature $T=J_{1}/100$~\cite{footnote}. The solid (dashed) hexagons depict the first (extended) Brillouin zones of the kagome lattice. The momenta $(k_{x},k_{y})$ are expressed in units where the edge length of the kagome triangles is set to unity.
}\label{Fig:PhaseDiagram}
\end{figure}

We now proceed to determine the in-plane exchange couplings for CaCu$_3$(OH)$_6$Cl$_2\cdot0.6$H$_2$O by mapping the energies for selected spin configurations on the Heisenberg Hamiltonian.
In Table~\ref{Tab:DFT}, we list the $J_i$ for several values of the on-site interaction $U$, calculated for the $T=213$~K structure reported in Ref.~[\onlinecite{CaCu3_Yoshida_2017}].
Interestingly, the $T=5$~K structure yields almost identical Heisenberg Hamiltonian parameters; this is surprising as exchange interactions can be quite temperature dependent even without change of space group symmetry.
The Curie-Weiss temperature for $J$s with $U=8$~eV is in good agreement with the in-plane Curie-Weiss temperature $\Theta_\text{CW}=-56.5$~K determined by the magnetic susceptibility measurements~\cite{CaCu3_Yoshida_2017}.
The magnetic susceptibilities are calculated by using the $J$s obtained by our DFT calculations, which are compared to the experimental results as shown in Fig.~\ref{Fig:PhaseDiagram}(a).
The set of $J$s for $U=8.0$~eV excellently reproduce the magnetic susceptibilities of CaCu$_3$(OH)$_6$Cl$_2\cdot0.6$H$_2$O, confirming the validity of our DFT calculations.
We also fit the magnetic susceptibilities by HTSE using the DFT results as the initial parameters, yielding $J_1=52.6$~K, $J_2=13.7$~K, and $J_d=-1.29$~K~\cite{SM}.
Both DFT calculations and the HTSE fitting~\cite{CaCu3_Yoshida_2017,SM} indicate the presence of sizable antiferromagnetic $J_2$ and ferromagnetic $J_d$.
Therefore, our results demonstrate that centennialite is the first $J_1$-$J_2$-$J_d$ kagome lattice antiferromagnet and provide the fertile playground for geometrical frustration.

The classical phase diagram of the $J_1$-$J_2$-$J_d$ kagome-lattice magnet with antiferromagnetic $J_1$~\cite{ClassicalEnergy,HTSE} is illustrated in Fig.~\ref{Fig:PhaseDiagram}(b).
Both sets of $J$s in CaCu$_3$(OH)$_6$Cl$_2\cdot0.6$H$_2$O determined by our DFT calculations and the HTSE fitting are located in the $\mathbf{q}=\mathbf{0}$ phase of the classical phase diagram, however, given the spin-$\frac{1}{2}$ nature of the system and its proximity to a tricritical point, one expects amplified quantum fluctuations with the low-energy physics being governed by a subtle interplay of quantum corrections to the different magnetic orders. Indeed, our PFFRG analysis for the DFT interaction parameters reveals the presence of a quantum paramagnetic phase which displays magnetic fluctuation tendencies towards $\mathbf{q}=\mathbf{0}$ order as confirmed by the presence of soft maxima at the expected location [$\mathbf{q}=(0,2\pi/\sqrt{3}$) and symmetry related positions, i.e., the center of the sides of the extended Brillouin zone] of the incipient Bragg peaks [see Fig.~\ref{Fig:PhaseDiagram}(c)] seen in neutron diffraction measurements [Fig.~\ref{Fig:Diffraction}(b)]. Ultimately, the onset of $\mathbf{q}=\mathbf{0}$ order is triggered by a nonnegligible out-of-plane DM interaction ($D_{z}>0$) whose magnitude in centennialite, as reported by thermal conductivity measurements~\cite{CaCu3_Doki_2018}, is estimated to be $D_{z}/J_{1}\sim0.1$. Given the fact, that in the $J_{1}$ only model the kagome spin liquid gives way to $\mathbf{q}=\mathbf{0}$ order for $D_{z}/J_{1}\sim0.1$~\cite{DM_Cepas_2008,Hering-2017,Buessen-2019}, one may expect that owing to the presence of small antiferromagnetic $J_{2}$ ($4\%$ of $J_{1}$) and ferromagnetic $J_{d}$ ($\sim2\%$ of $J_{1}$) both of which favor $\mathbf{q}=\mathbf{0}$ spin pattern~\cite{Suttner-2014,Iqbal-2015}, that a slightly smaller value $D_{z}$ might suffice to drive the system out of the quantum paramagnetic phase and induce $\mathbf{q}=\mathbf{0}$ long-range magnetic order, as seen in experiments. These results indicate that while centennialite is located in the $\mathbf{q}=\mathbf{0}$ ordered phase, it is not immersed deep inside it, but rather precariously placed in the vicinity of a quantum critical point. This also explains the reduced magnetic moment in centennialite.
%We find that the $J$s of CaCu$_3$(OH)$_6$Cl$_2\cdot0.6$H$_2$O are close to the phase boundaries and hence probably in the vicinity of a quantum critical point considering the quantum phase diagram for ferromagnetic $J_1$~\cite{MgKape_Boldrin_2015}.
In fact, low temperature heat capacity measurements revealed the existence of the $T$ linear term in addition to the spin wave term~\cite{CaCu3_Yoshida_2017}. The $T$ linear term suggests that the continuum excitations exist even below $T_\text{N}$. On the other hand, spinon like behavior is also observed above $T_\text{N}$ in thermal conductivity measurements of centennialite~\cite{CaCu3_Doki_2018}.

%Such spinon behavior can be investigated more directly by inelastic neutron scattering technique, which is under way.
%In addition, it is also intriguing to theoretically figure out the quantum phase diagram of the $J_1$-$J_2$-$J_d$ kagome antiferromagnet.
%These ingredients can help us understand the nature of the spin-1/2 $J_1$-only kagome lattice antiferromagnet.

In summary, we investigate in detail the crystal and magnetic structures of CaCu$_3$(OH)$_6$Cl$_2\cdot0.6$H$_2$O.
Cu$^{2+}$ ions form a perfect kagome lattice without intersite disorder between Cu$^{2+}$ and Ca$^{2+}$.
Magnetic diffraction measurements determine the $\mathbf{q}=\mathbf{0}$ magnetic structure with the negative vector spin chirality.
The magnetic propagation vector is $\mathbf{k}=(0,0,0.5)$ and the ordered moment at 0.3~K is suppressed to be $0.58(2)\mu_\text{B}$.
Exchange couplings of CaCu$_3$(OH)$_6$Cl$_2\cdot0.6$H$_2$O are estimated by a DFT calculations to be $J_1=55$~K, $J_2=2.2$~K, and $J_d=-1.7$~K, and a spin-$\frac{1}{2}$ PFFRG analysis of the same shows the presence of magnetic fluctuation tendencies towards $\mathbf{q}=\mathbf{0}$ magnetic structure. The estimated DM interaction of $D_{z}\sim0.10$ in centennialite likely places the system on the edge of a phase transition but in the $\mathbf{q}=\mathbf{0}$ ordered phase. The present results demonstrate that CaCu$_3$(OH)$_6$Cl$_2\cdot0.6$H$_2$O is a first realization of a $J_1$-$J_2$-$J_d$ kagome-lattice magnet with antiferromagnetic $J_1$ and is located in the vicinity of a quantum critical point.

{\it Acknowledgments}. We thank Yoshihiko Ihara and Minoru Yamashita for helpful discussions.
We also thank for technical support from the MLF sample environment team.
Single crystals were checked at the CROSS user laboratories.
Synchrotron x-ray diffraction at PF was carried out under the proposal number 2017S2-001.
Neutron experiments at MLF were conducted under the user program with the proposal numbers 2017A0013 and 2017BU1401 for AMATERAS, 2017I0001 for 4SEASONS, and 2018A0009 and 2018B0008 for SENJU.
Present work was supported by JSPS KAKENHI Grant Number JP18K03529 and the Cooperative Research Program of ``Network Joint Research Center for Materials and Devices'' (20181072). Y.I. acknowledges the Science and Engineering Research Board (SERB), DST, India for support through Startup Research Grant No. SRG/2019/000056 and MATRICS project No. MTR/2019/001042. This research was supported in part by the International Centre for Theoretical Sciences (ICTS) during a visit for participating in the program -  Novel phases of quantum matter (Code: ICTS/topmatter2019/12) and in the program - “The 2nd Asia Pacific Workshop on Quantum Magnetism” (Code: ICTS/apfm2018/11). This research was supported in part by the National Science Foundation under Grant No. NSF PHY-1748958.


\begin{thebibliography}{apsrev4-2}
\bibitem{footnote}
The one-loop PFFRG calculations are done with the longest spin-spin correlator being 16 lattice spacings and a discretization over 64 frequencies of the vertex functions.
\end{thebibliography}
\end{document}